\documentclass[12pt]{article}
\usepackage{epsfig}
\usepackage{amsmath}
\usepackage{graphicx}
\usepackage{latexsym}
\usepackage{subfigure}
\begin{document}

\begin{center}
{\bf \large STABILITY OF YANG-MILLS FIELDS SYSTEM\\ IN THE HOMOGENEOUS (ANTI-)SELF-DUAL BACKGROUND FIELD}\\
\vspace{0.5cm} {\it \rmfamily \large V.I. Kuvshinov\footnote{E-mail:
V.Kuvshinov@sosny.bas-net.by},
V.A. Piatrou\footnote{E-mail: PiatrouVadzim@tut.by}}\\
Joint Institute for Power and Nuclear Research, Minsk, Belarus
\end{center}

\begin{abstract}
Stability of Yang-Mills fields system in the background field is
investigated basing on Toda criterion, Poincare sections and the
values of the maximal Lyapunov exponents. The existence of the
region of regular motion at low densities of energy is demonstrated.
Critical energy density of the order-chaos transition is analyzed
for the different values of the model parameter.
\end{abstract}

\section*{Introduction}

In contrast to electrodynamics, the dynamics of Yang-Mills fields is
inherently nonlinear and chaotic at any density of energy. This
assumption was confirmed analytically and numerically
\cite{81MatSavTer, 81ChirShep, 83Sav}. Further analysis of spatially 
homogeneous field configurations \cite{79BasMatSav} showed that
inclusion of Higgs field leads to order-chaos transition at some
density of energy of classical gauge fields \cite{81MatSavTerL,
84Sav, 85BerManSad}. Classical Higgs field regularizes chaotic
dynamics of classical gauge fields below critical energy density and
leads to the emergence of order-chaos transition.

Chaos in Yang-Mills fields \cite{02KKPL} and vacuum state
instability in nonperturbative QCD models \cite{77Sav, 78NieOle,
79NieOle} are also considered in connection with confinement. It has
been also shown recently that interaction of the constant
chromo-magnetic field with axial field could generate confinement
\cite{06GaeSpa}. These results indicate the importance of
nonperturbative background fields.

In our previous paper \cite{05GK} we have investigated the stability
of Yang-Mills-Higgs fields and described analytically the regions of
chaotic and stable motion. In this work Yang-Mills fields are
considered on the background of the homogeneous (anti-) self-dual
field \cite{77BatMatSav}. As the dynamics of arbitrary field
configurations is too complicated, we follow \cite{82ChirShep} and
reduce our model to spatially homogeneous fields which depend on
time. After that we are left with only finite number of degrees of
freedom (two in our case) which allows us to investigate the
dynamics of the system using conventional methods developed for
mechanical systems.

In this work one more mechanism of the chaos suppression in the
Yang-Mills fields models is proposed. Homogeneous (anti-)self-dual
field eliminates chaoticity of Yang-Mills dynamics below critical
energy density.

\section{Homogeneous (anti-)self-dual field}

In this paper classical dynamics of SU(2) model gauge fields system
is considered on the background of the homogeneous (anti-)self-dual
field. Various properties of this solution of the Yang-Mills
equations in SU(2) theory were investigated originally by other
authors \cite{77Sav, 80Leu, 81Min, 81Leu}. It was demonstrated that
self-dual homogeneous field provides the Wilson confinement
criterion \cite{99EfKalNed}. Therefore this field is at least a
possible source of confinement in QCD if it is a dominant
configuration in the QCD functional integral.

Homogeneous self-dual field is defined by the following expressions
\cite{77BatMatSav}:
$$
B^{a}_{\mu}=B n^{a} b_{\mu \nu} x_{\nu},
$$
$$
F^{a}_{\mu \nu}=- 2 B n^{a} b_{\mu \nu},
$$
where $B$ - value of the field strength, vector $n^{a}$ and tensor
$b_{\mu \nu}$ characterize the direction of the field, respectively,
in color space and in space-time. The latter has the following
properties
$$
b_{\mu \nu}=-b_{\nu \mu}, \qquad b_{\mu \nu} b_{\mu \rho}=
\delta_{\nu \rho},
$$
$$
\widetilde{b}_{\mu \nu}=\frac 1 2 \varepsilon_{\mu \nu \alpha \beta}
b_{\alpha \beta}=\pm b_{\mu \nu},
$$
where positive and negative signs in last expression correspond,
respectively, to self-dual and anti-self-dual cases.

As the directions of the background field in color space and in
space-time can be chosen arbitrarily, we will assume that the gauge
field has color components $(n^{1},n^{2},n^{3}) =(0,0,1)$ and
space-time components \( \textbf{B}=(B_{1},B_{2},B_{3})=(0,0,B)\).

\section{Model potential of the system}

The Lagrangian of SU(2) gauge theory in Euclidean metrics is
$$
L=-\frac{1}{4} G^{a}_{\mu \nu} G^{a}_{\mu \nu},
$$
where $G^{a}_{\mu \nu}$ is a field tensor which has the following
form:
$$
\label{Strength}
 G^{a}_{\mu \nu}=\partial_{\mu} A^{a}_{\nu} -
\partial_{\nu} A^{a}_{\mu} +
g \epsilon^{a b c} A^{b}_{\mu} A^{c}_{\nu}.
$$
In last expression  $A^{a}_{\mu}, a=1,2,3$ are the three non-abelian
Yang-Mills fields and  $g$ denotes the coupling constant of these
fields.

We consider the fluctuations around background homogeneous self-dual
field. Self-dual field is regarded as external one and it is taken
into account by substituting modified vector potential in the
Yang-Mills Lagrangian
$$
A^{a}_{\mu} \rightarrow A^{a}_{\mu}+B^{a}_{\mu}
$$
where $A^{a}_{\mu}$ is the fluctuation to the background field
$B^{a}_{\mu}$ .

We use the gauge:
$$
A^{a}_{4}= 0,
$$
and consider spatially homogeneous field configurations
\cite{82ChirShep}
$$
\partial_{i}A^{a}_{\mu}= 0, \qquad
i=1..3.
$$

Our model of Yang-Mills fields in (anti-)self-dual field is
constructed in Euclidean space. In order to analyze the model by
using analytical and numerical methods we should switch to Minkowski
space. We consider chromo-magnetic model. Thus we put
chromo-electric field is equal to zero. If $A_{1}^{1}=q_{1}$,
$A_{2}^{2}=q_{2}$ and the other components of the perturbative
Yang-Mills fields are equal to zero, the potential of the model is:
$$
V = \frac12 g^{2} q^2_1 q^2_2 + \frac12 H^{2} - g H q_1 q_2 +
\frac18 g^2 H^2 ( x^2 q_1^2 + y^2 q_2^2),\eqno{(1)}
$$
where $H$ - chromo-magnetic background field strength, $x$ and $y$ -
coordinates which play the role of the parameters,  $q_1$ and $q_2$
- field variables.

\section{Stability of the model}

\subsection{Toda criterion}

\begin{figure}[ht!]
\epsfig{file = 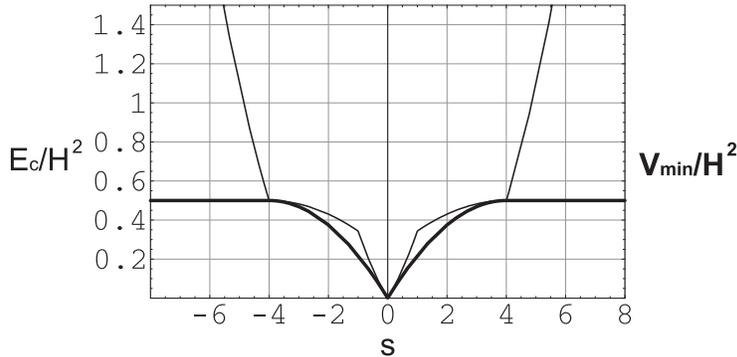, height = 0.2\textheight} \centering
\leavevmode \caption{Critical energy density of order-chaos
transition (thin line) and minimum of the energy (thick line) as a
functions of the model parameter $s = g H x y$.}\label{GrEcVmin}
\end{figure}

At first, stability of the model is investigated using well known
technique based on the Toda criterion of local instability
\cite{74Toda, Salasnich} which allows us to obtain the value of the
critical energy density of order-chaos transition in the system.
This energy and minimum of the energy as the functions of the model
parameter $s=g H x y$ are shown on the fig.\ref{GrEcVmin}.

Critical and minimal energies are close to each other for $s\in(-4,
4)$. This behavior indicates the absence of the region of regular
motion in the system. In other case $\left(s\in(-\infty, -4) \mbox{
or } s\in(4, \infty)\right)$, the critical energy density is much
larger than minimal one and the system is regular up to this energy.
These results will be checked using numerical methods in next
subsection.

\subsection{Numerical calculations}

The system is investigated using Poincare sections and Lyapunov
exponents for wide range of model parameter values. These numerical
methods could indicate global regular regimes of motion whereas Toda
criterion reveals only the local chaotic properties of the
trajectories \cite{77BenBraGal}. Thus numerical methods are more
precise for stability analysis.
\begin{figure}[ht!]
\center
\subfigure[$E = 0.005$]{\includegraphics[angle = 270,width = 0.45\textwidth]{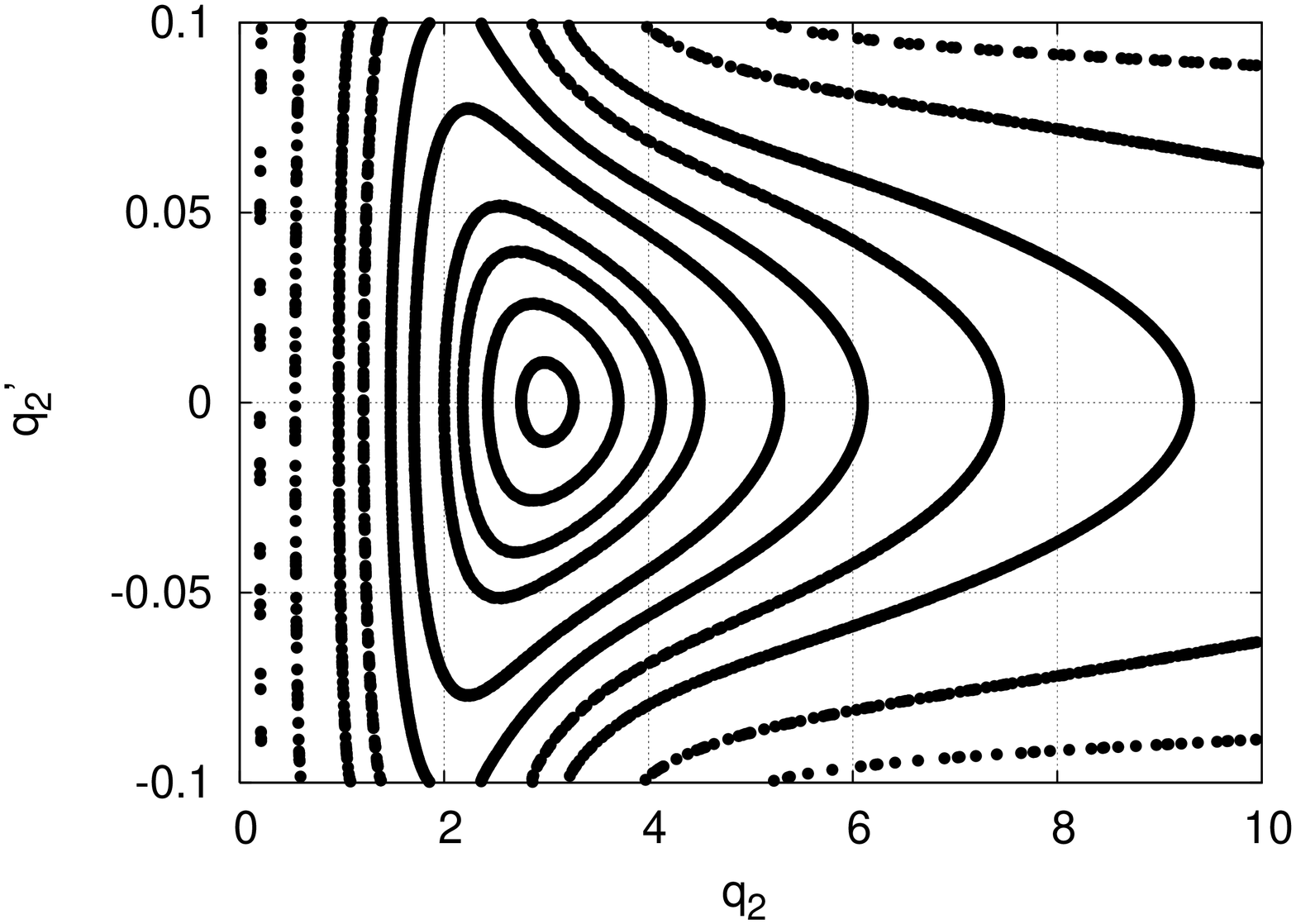}}\hfil
\subfigure[$E =0.15$]{\includegraphics[angle = 270,width = 0.45\textwidth]{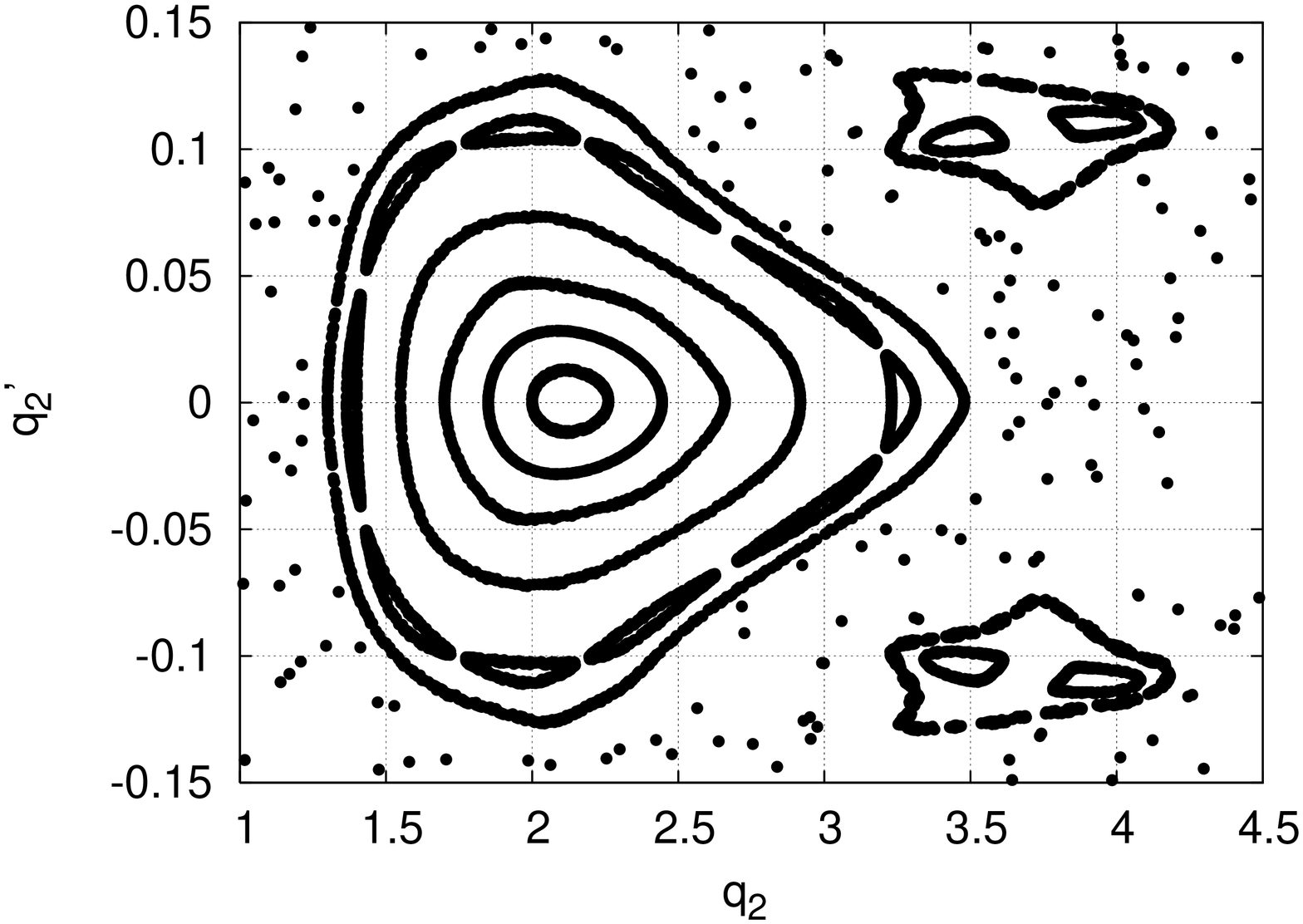}}
\caption{Poincare sections for two dimensional Yang-Mills system in
the background field for $s = 0, H = 1$.}\label{Grps0}
\end{figure}

\begin{figure}[ht!]
\epsfig{file = 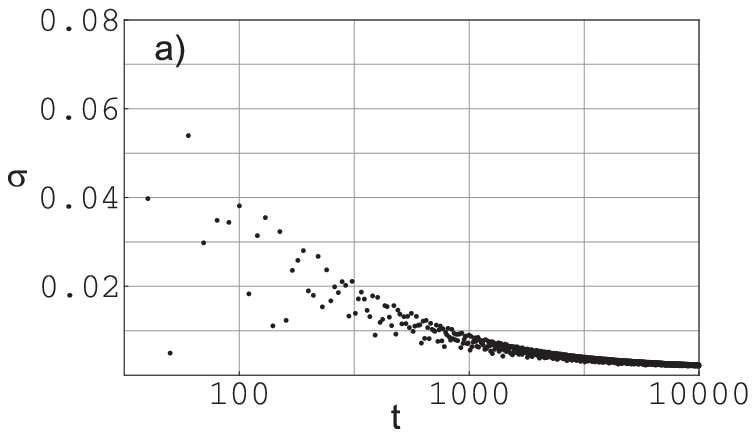, width = 0.45\textwidth}
\epsfig{file = 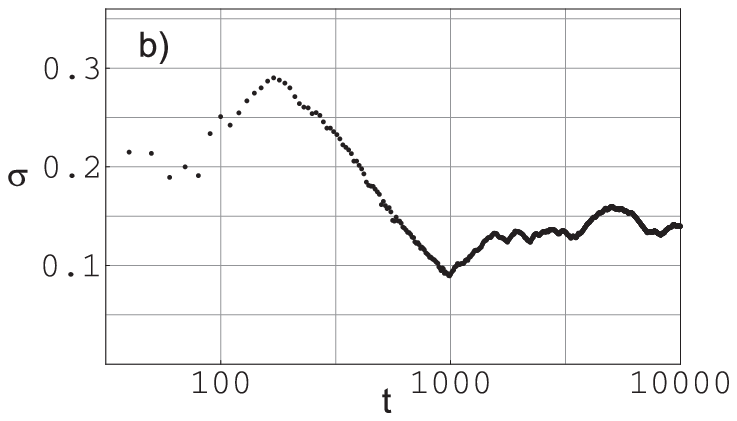, width = 0.45\textwidth}
\caption{Maximal Lyapunov exponents for two dimensional Yang-Mills
system in the background field for $s = 0, H = 1,(a) E = 0.15$ and
$(b) E = 0.68$.}\label{Grle0}
\end{figure}

Results of the numerical calculations for the system with model
parameter $s = 0$ are shown on the fig.\ref{Grps0} and
fig.\ref{Grle0}. Contrary to Toda criterion, system is regular at
small energies below the energy of the background field $E_c =
E_{vac} = \frac12 H^2$. There are only regular regimes of motion for
small energies (e.g. fig.\ref{Grps0}a and fig.\ref{Grle0}a) and two
types of motion for energies above $E_c$ (fig.\ref{Grps0}b and
fig.\ref{Grle0}b).

\begin{figure}[ht!]
\center \subfigure[$E = 21$]{\includegraphics[angle = 270,width =
0.45\textwidth]{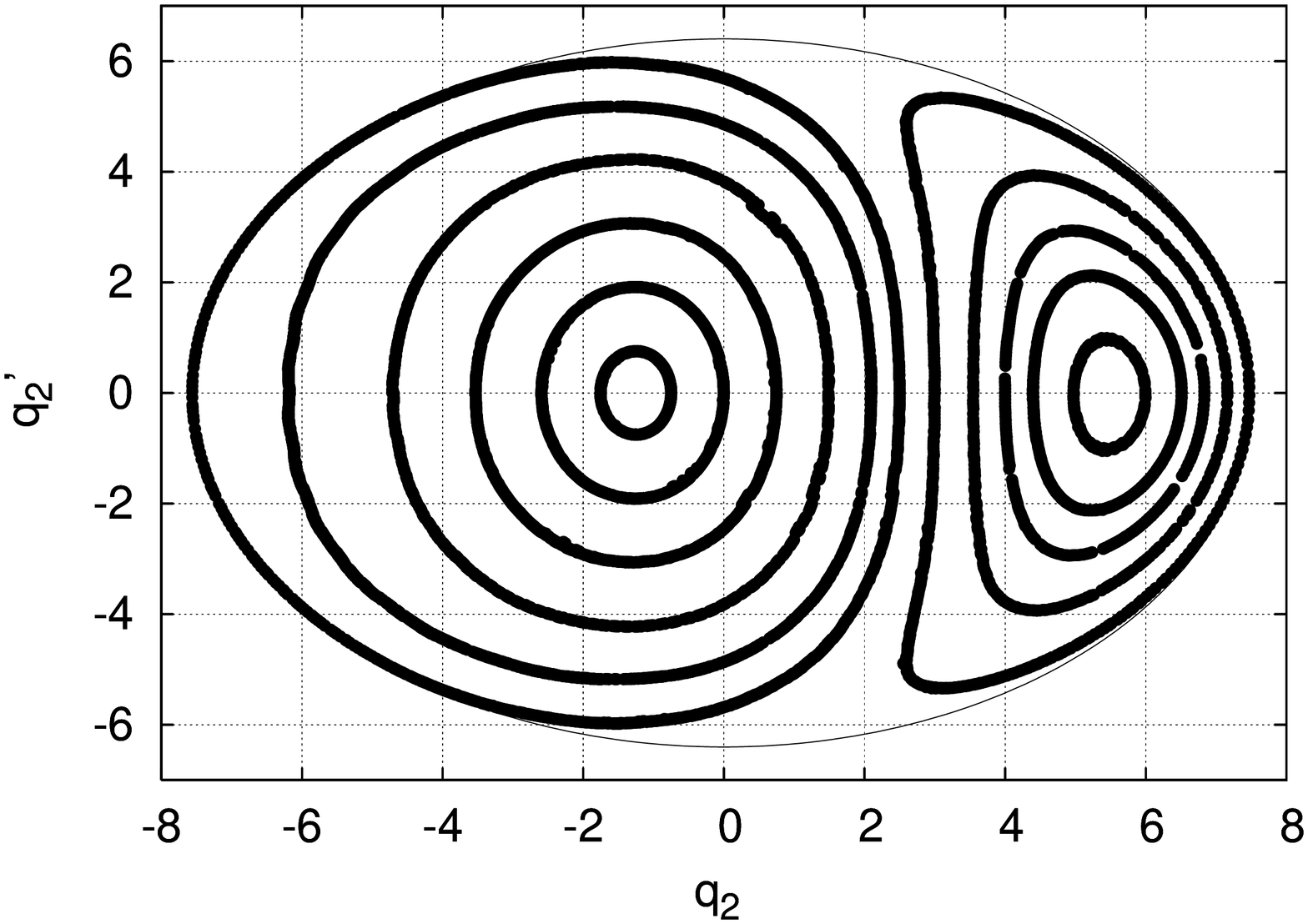}}\hfil \subfigure[$E =
100$]{\includegraphics[angle = 270,width =
0.45\textwidth]{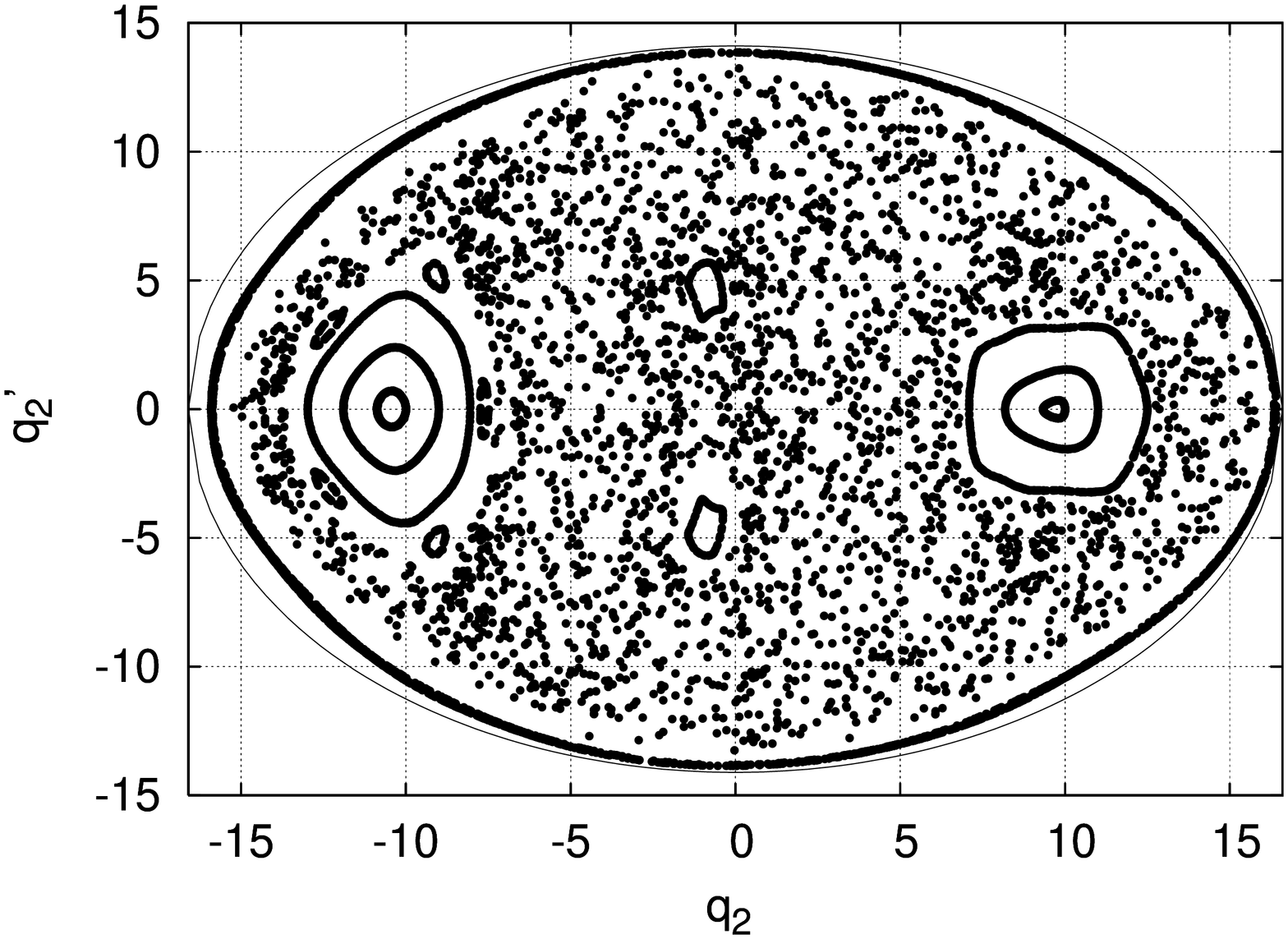}} \caption{Poincare sections for two
dimensional Yang-Mills system in the background field for $s = 25.5,
g = 0.1, H = 1, x = 15, y = 17, (a) E = 21$ and $(b) E = 100$. Thin
line - border of the phase space.}\label{Grps25}
\end{figure}

We have the following Poincare sections (fig.\ref{Grps25}) for large
model parameter values. It is seen that system is fully regular
(fig.\ref{Grps25}a) for high values of energy $(E_{vac} << E < E_c)$
as it was revealed by Toda criterion. All trajectories have zero
maximal Lyapunov exponents for this case. Two types of trajectories
(chaotic and regular) are present in the system with energies above
the critical one (fig.\ref{Grps25}b).

It is seen that Toda criterion rather good describes the region of
the large values of model parameter $s$ and fails for $s\in(-4,4)$.

Numerical calculations have shown that there is a region of regular
motion at low densities of energy in our system at any value of the
model parameter. Therefore, homogeneous (anti-)self-dual field
regularizes chaotic dynamics of Yang-Mills fields system.

\section*{Conclusions}

The existence of the nonperturbative component of the Yang-Mills
field is crucial for the confinement phenomenon. On the other hand,
classical dynamics of Yang-Mills field is chaotic at any density of
energy in the absence of background fields and can be regularized
only if some other fields, for example, the Higgs field, are
included in the model.

In this work we have demonstrated that homogeneous (anti-)self-dual
background field has similar properties. Yang-Mills field on such
background has the region of regular motion at low densities of
energy. There is order-chaos transition in the system at any values
of model parameters. The critical density of energy of this
transition is equal to the background field energy for small
parameters and is much larger for large parameter values.

\section*{Acknowledgements}

This work is supported by Belarusian Republican Foundation for
Fundamental Research.


\begin{thebibliography}{10}

\bibitem{81MatSavTer}
\textit{Matinyan~S.~G., Savvidy~G.~K., Ter-Arutunyan-Savvidy~N.~G.}
Classical Yang-Mills mechanics. Nonlinear color
oscilations~//{Sov.Phys.JETP.} 1981. V.~53. P.~421.

\bibitem{81ChirShep}
\textit{Chirikov~B.~V., Shepelyansky~D.~L.} Stochastic oscillations
of classical Yang-Mills fields~//{JETP Lett.} 1981. V.~34, No.~4.
P.~163--166.

\bibitem{83Sav}
\textit{Savvidy~G.~K.} The Yang-Mills classical mechanics as a
Kolmogorov K-system~//{Phys. Lett.} 1983. V. B130, No.~5.
P.~303--307.

\bibitem{79BasMatSav}
\textit{Baseyan~G.~Z., Matinyan~S.~G., Savvidy~G.~K.} Nonlinear
plane waves in massless Yang-Mills theory~//{Pis'ma Zh. Eksp. Teor.
Fiz.} 1979. V.~29, No.~10. P.~641--644.

\bibitem{81MatSavTerL}
\textit{Matinian~S.~G., Savvidy~G.~K., Ter-Arutunian~N.~G.}
Stochastisity of Yang-Mills mechanics and its elimination by Higgs
mechanism~//{Pisma Zh.Eksp.Teor.Phys.} 1981. V.~34, No.~11.
P.~613--616.

\bibitem{84Sav}
\textit{Savvidy~G.} Classical and quantum mechanics of nonabelian
gauge fields~//{Nucl.Phys.} 1984. V. B246. P.~302--359.

\bibitem{85BerManSad}
\textit{Berman~G.~P., Mankov~Y.~I., Sadreyev~A.~F.} Stochastic
instability of classical homogeneous SU(2)$\bigotimes$U(1) fields
with spontaneously broken symmetry~//{Zh. Eksp. Teor. Fiz. (JETP)}.
1985. V.~88, No.~3. P.~705--714.

\bibitem{02KKPL}
\textit{Kuvshinov~V.~I., Kuzmin~A.~V.} Towards chaos criterion in
quantum field theory~//{Phys. Lett. A.} 2002. V. 296. P.~82--86, hep-th/0205263.

\bibitem{77Sav}
\textit{Savvidy~G.~K.} Infrared instability of the vacuum state of
gauge theories and asymptotic freedom~//{Phys.Lett.} 1977. V. 71B.
P.~133--134.

\bibitem{78NieOle}
\textit{Nielsen~N.~K., Olesen~P.} An unstable Yang-Mills field
mode~// {Nucl. Phys.} 1978. V.~B144. P.~376--396.

\bibitem{79NieOle}
\textit{Ambj{\o}rn~J., Nielsen~N.~K., Olesen~P.} A hidden Higgs
lagrangian in QCD~//{Nucl. Phys.} 1979. V. B152. P.~75--96.

\bibitem{06GaeSpa}
\textit{Gaete~P., Spallucci~E.} Confinement effects from interacting
chromo-magnetic and axion fields~//{J. Phys. A}. 2006. V.~39.
P.~6021--6030, hep-th/0512178.

\bibitem{05GK}
\textit{Kuvshinov~V.~I., Piatrou~V.~A.} Stability of
Yang-Mills-Higgs field system in the homogeneous self-dual vacuum
field~// {Nonlin.Phenom.Complex Syst.} 2005. V.~8, No.~2.
P.~200--205, nlin.CD/0511060.

\bibitem{77BatMatSav}
\textit{Batalin~I.~A., Matinian~S.~G., Savvidy~G.~K.} Vacuum
polarization by a source-free gauge field~//{Sov. J. Nucl. Phys.}
1977. V.~26. P.~214.

\bibitem{82ChirShep}
\textit{Chirikov~B.~V., Shepelyansky~D.~L.} Dynamics of some
homogeneous models of classical Yang-Mills fields~//{Sov. J.Nucl.
Fiz.} 1982. V.~36, No.~6. P.~908--915.

\bibitem{80Leu}
\textit{Leutwyler~H.} Vacuum fluctuations surrounding soft gluon
fields~//{Phys.Lett.} 1980. V. 96B. P.~154--158.

\bibitem{81Min}
\textit{Minkowski~P.} On the ground-state expectation value of the
field strength bilinear in gauge theories and constant classical
fields~//{Nucl.Phys.} 1981. V. B177. P.~203--217.

\bibitem{81Leu}
\textit{Leutwyler~H.} Constant gauge fields and their quantum
fluctuations~//{Nucl.Phys.} 1981. V. B179. P.~129--170,.

\bibitem{99EfKalNed}
\textit{Efimov~G.~V., Kalloniatis~A.~C., Nedelko~S.~N.} Confining
properties of the homogeneous self-dual field and the effective
potential in SU(2) Yang-Mills theory~//{Phys.Rev.} 1999. V. D59.
P.~014026, hep-th/9806165.

\bibitem{74Toda}
\textit{Toda~M.} Instability of trajectories of the lattice with
cubic nonlinearity~//{Phys.Lett.} 1974. V. A 48, No.~5. P.~335--336.

\bibitem{Salasnich}
\textit{Salasnich~L.} Quantum signature of the chaos-order
transition in a homogeneous SU(2) Yang-Mills-Higgs
system~//{Phys.Atom.Nucl.} 1998. V.~61. P.~1878--1881.

\bibitem{77BenBraGal}
\textit{Benettin~G., Brambilla~R., Galgani~L.} A comment on the
reliability of the Toda criterion for the existence of a stochastic
transition~// {Physica}. 1977. V. A87, No.~2. P.~381--390.

\end{thebibliography}
\end{document}